\documentclass[aps,prl,reprint,groupedaddress,amsmath,amssymb]{revtex4-1}
\usepackage{amsmath}
\usepackage{graphicx}
\usepackage{stackengine}
\usepackage{color}
\usepackage[T1]{fontenc}
\begin{document}
	
	\title{Curvature variation controls particle aggregation on fluid vesicles}
	\author{Afshin Vahid$^1$, An\dj ela \v{S}ari\'{c}$^2$, Timon Idema$^1$}
	\email[]{T.Idema@TUDelft.nl}
	\affiliation{\small \em $^1$\textit{Department of Bionanoscience, Kavli Institute of Nanoscience, Delft University of Technology, Delft, The Netherlands}}
	\affiliation{\small \em
		$^2$ \textit{Department of Physics and Astronomy, Institute for the Physics of Living Systems, University College London,London, United Kingdom}}

 	\date{\today}
	
	\begin{abstract}
		Cellular membranes exhibit a large variety of shapes, strongly coupled to their function. Many biological processes involve dynamic reshaping of membranes, usually mediated by proteins. This interaction works both ways: while proteins influence the membrane shape, the membrane shape affects the interactions between the proteins. To study these membrane-mediated interactions on closed and anisotropically curved membranes, we use colloids adhered to ellipsoidal membrane vesicles as a model system. We find that two particles on a closed system always attract each other, and tend to align with the direction of largest curvature. Multiple particles form arcs, or, at large enough numbers, a complete ring surrounding the vesicle in its equatorial plane. The resulting vesicle shape resembles a snowman. Our results indicate that these physical interactions on membranes with anisotropic shapes can be exploited by cells to drive macromolecules to preferred regions of cellular or intracellular membranes, and utilized to initiate dynamic processes such as cell division. The same principle could be used to find the midplane of an artificial vesicle, as a first step towards dividing it into two equal parts.
	\end{abstract}
	
	\pacs{}
	
	\maketitle
	\section{Introduction}
		
	
	\begin{figure*}[ht]
		\centering
		\includegraphics[width=0.48\textwidth]{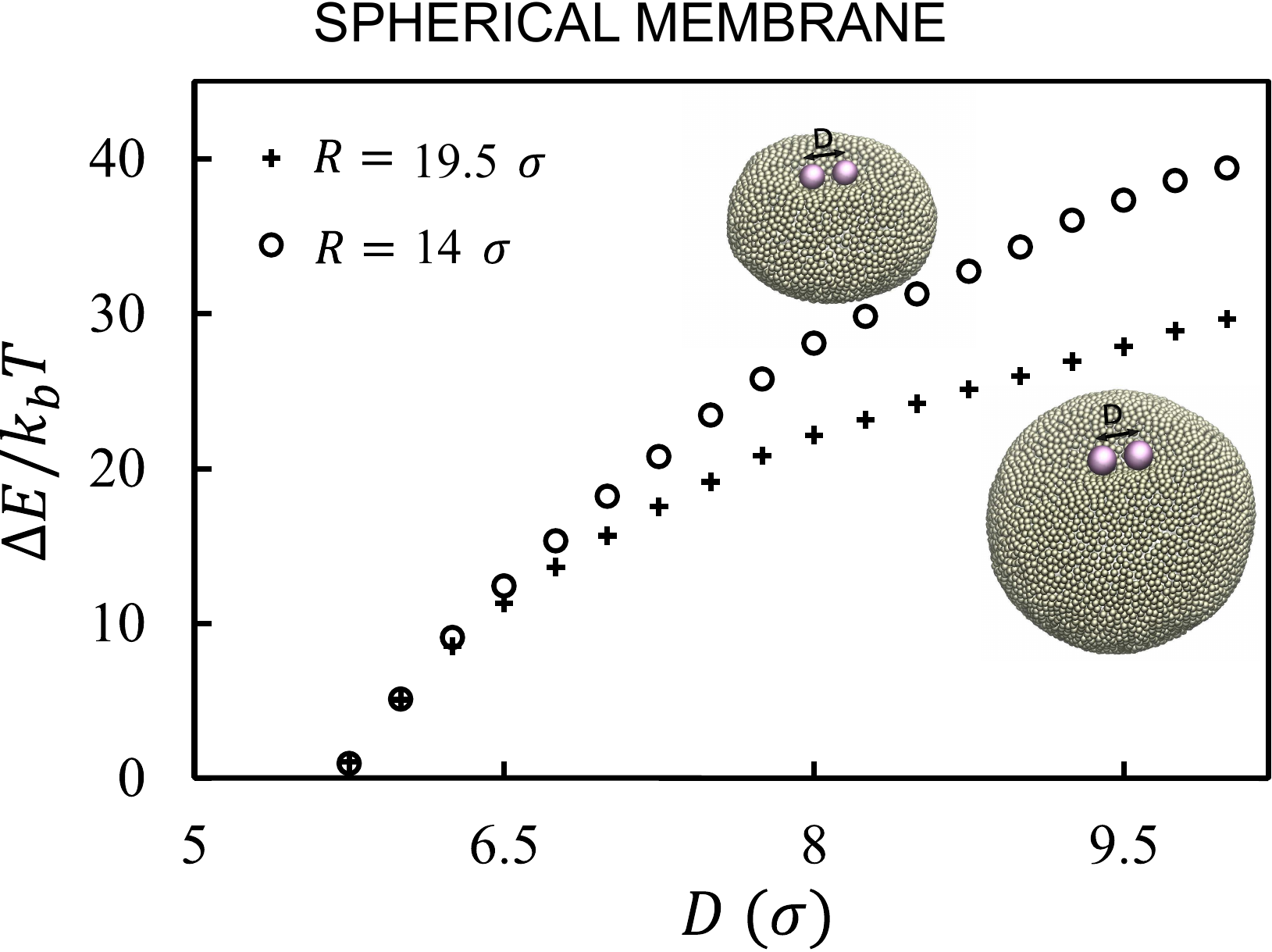}
		\caption{The curvature energy of the membrane for spherical vesicles of sizes $D_v = 14 \sigma$ (circles) and  $D_v = 19.5 \sigma$ (pluses) containing two colloids. The force between colloids in the smaller vesicle is stronger and has a larger interaction range.}
		\label{fig:TwoVesicles}
	\end{figure*}
	Cellular membranes are two-dimensional fluid interfaces that consist of a large variety of components. They form the boundary between the cell and the outside world, and, for eukaryotic cells, separate the inside of the cell into numerous compartments known as organelles. In order for biological processes like cell division, vesicular trafficking and endo/exocytosis to occur, cellular membranes have to reshape constantly. Consequently, membranes exhibit a variety of morphologies, from a simple spherical liposome to bewildering complex structures like interconnected tubular networks as found in Mitochondria and the Endoplasmic Reticulum (ER), or connected stacks of perforated membrane sheets in the Golgi apparatus\citealp{mcmahon2015membrane,schmick2014interdependence,shibata2009mechanisms,mcmahon2005membrane}. There are different mechanisms by which membranes achieve these structures, the most important of which is through the interplay between membrane lipids and various proteins\cite{wang2016cooperation,johannes2014bending,drin2010amphipathic}. A biological membrane is home to different types of proteins that are adhered to or embedded in it. These proteins deform the membrane and, consequently, they can either repel or attract each other \cite{reynwar2007aggregation,dommersnes2002many,schweitzer2015membrane,kim1999many,weikl1998interaction,goulian1993long}. Spatial organization of such proteins in biological membranes is essential for stabilizing the membrane and for the dynamic behaviour of cellular organelles \cite{thalmeier2016geometry,english2013endoplasmic,pannuzzo2013alpha,ehrlich2004endocytosis}. 
	
	Recently, it has been experimentally \cite{PhysRevE.94.012604,peter2004bar} and theoretically \cite{kim1999many,Vahid2016Point,bahrami2013orientational,dommersnes2002many,reynwar2007aggregation,simunovic2013linear,vsaric2012fluid,vsaric2011particle,pamies2011reshaping} revealed that membrane-curving particles, like colloids or identical proteins, adhered to a membrane self-assemble into striking patterns. For instance it has been shown that colloids adhered to a spherical membrane form linear aggregations  \cite{vsaric2012fluid}. In all of the studies to date, the global shape of the membrane is selected from one of three options: planar, spherical, or tubular. These global membrane shapes impose a homogeneous background curvature, which is considered to be conserved throughout the process under investigation. Outside factors changing the membrane have not yet been included in the study of membrane-mediated interactions. Membranes in cellular compartments such as in ER and the Golgi complex are however dynamic entities and possess peculiar shapes forming regions with high local curvature and regions with less curvature \cite{shibata2009mechanisms}. Forming and stabilizing such shape inhomogeneities is necessary for cellular functions like sensing and trafficking \cite{aimon2014membrane}. It is therefore warranted to investigate how the interactions between membrane inclusions are affected by anisotropies in the membrane curvature.	
	
	In this study, through a numerical experiment, we investigate the interactions between colloids adhered to a quasi-ellipsoidal membrane with a varying curvature. We also include all other factors from earlier studies such as surface tension, adhesion energy (required for colloids to adhere to the membrane) and constant volume effects. We use a dynamic triangulation network to model the membrane, and computationally minimize the total energy of the membrane via a Monte Carlo algorithm. Firstly, we show that the interaction between two colloids adhered to spherical vesicles is significantly affected by the vesicle curvature. Secondly, we demonstrate that linear aggregates of colloids exploit the curvature anisotropy and adjust their orientation to minimize the total energy on a quasi-ellipsoidal membrane. Using umbrella sampling, we further show that the total energy of the membrane favors two colloids to attract each other at the mid-plane of a prolate ellipsoid that is perpendicular to its major axis. Finally, we investigate how the various terms in the total energy of the membrane affect the strength of the interactions. Our results show that the variation in the membrane shape can play a crucial role in a variety of cellular functions that require macromolecular assembly or membrane remodeling.

	\section{Model}
	
	The conformation of a fluid membrane can be described as the shape minimizing the classical Helfrich energy functional \cite{helfrich1973elastic}:\\
	\begin{align}
	u_\mathrm{Curv} = \frac{\kappa}{2}\int_{A} (2H)^2 dA,
	\end{align}
	where $H$ is the mean curvature at any point on the surface of the membrane and geometrically is defined as the divergence of the normal vector to the surface, $H= -\frac{1}{2} \nabla \cdot \mathbf{n}$. In our computational scheme, we discretize the membrane by a triangulated network, whose triangles represent course-grained patches of the membrane \cite{nelson2004statistical,vsaric2010effective}. Using a discretized form of the Helfrich energy, we define the curvature energy as:
	\begin{align}
	u_\mathrm{Curv} = \kappa \sum_{<ij>}^{} 1- \textbf{n}_i \cdot \textbf{n}_j,
	\end{align}
		\begin{figure*}[ht]
			\centering
			\includegraphics[width=0.55\textwidth]{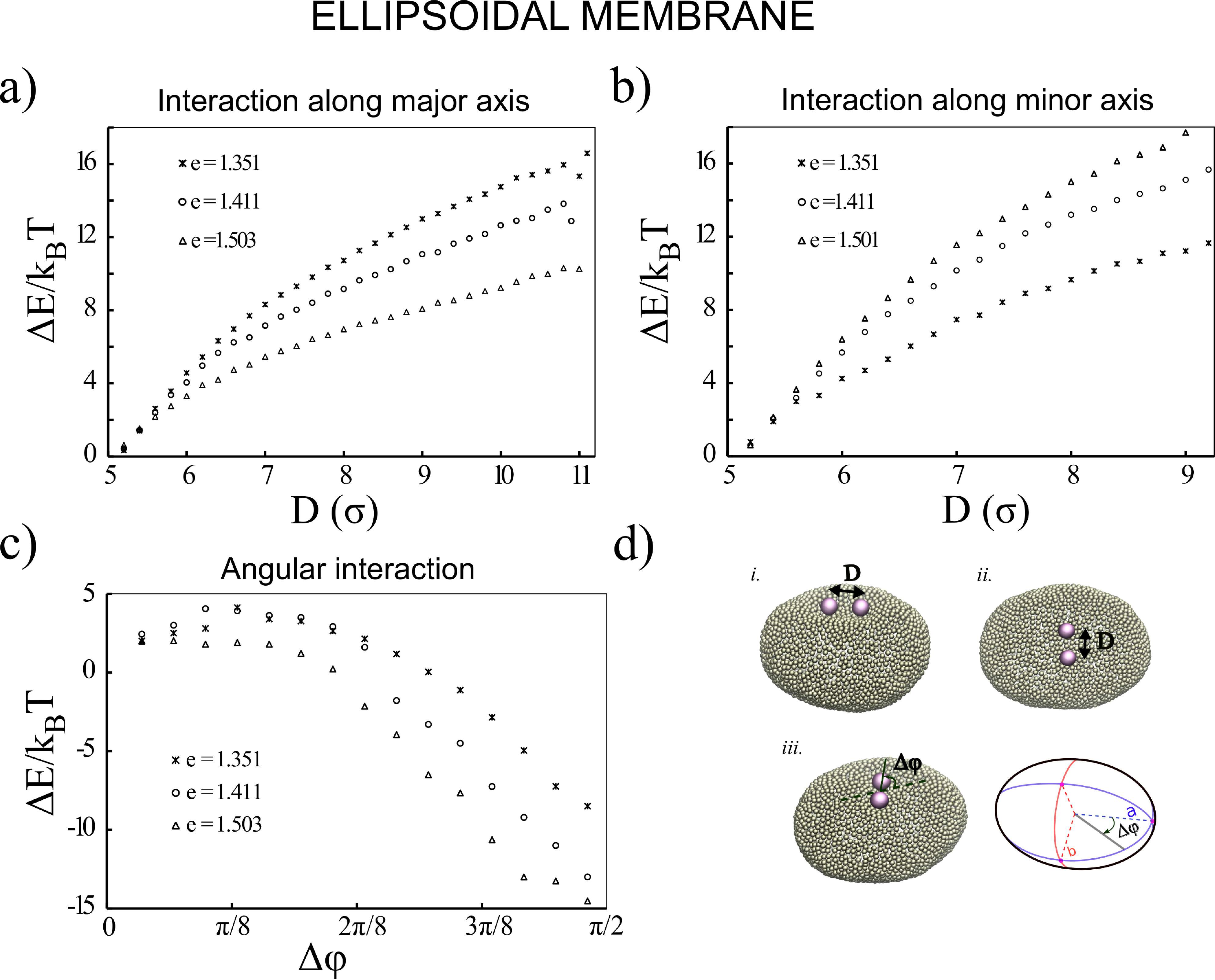}
			\caption{Colloids adhered to a quasi-ellipsoidal membrane behave differently in different directions. Decreasing the ellipticity of the vesicle, which is defined as $e=a/b$, (a) strengthens the attractions between the colloids along the major axis and (b) weakens the interaction between the colloids along the minor axis. Figure (c) illustrates that the energy of a membrane containing a pair of colloids decreases when the angle between the pair and the semi-major axis increases.}
			\label{fig:fig2}
		\end{figure*}
	where $\textbf{n}_i$ and $\textbf{n}_j$ are the normal vectors to any pair of adjacent triangles $i$ and $j$, respectively. The summation runs over all pairs of such triangles. In order to guarantee the fluidity of the membrane, we cut and reattach the connection between the four vertices (which we label with and refer to as beads) of any two neighboring triangles. The membrane in our system does not undergo any topological changes and we can thus ignore the Gaussian curvature contribution in the bending energy. We impose the conservation of membrane surface area ($A$) and enclosed volume ($V$) by adding the terms $u_{\mathrm{A}}= K_{\mathrm{A}} (A-A_t)^2/A_t$ and $u_{\mathrm{V}}= K_{\mathrm{V}} (V-V_t)^2/V_t$ to the energy during the minimization process, with $A_t$ and $V_t$ the target values of the membrane's area and enclosed volume. The corresponding constants are chosen such that both the area and volume deviate less than $0.05 \% $ from their target values. To enable colloids to adhere to the membrane, we introduce an adhesion potential, $u_{\mathrm{Ad}} = -\varepsilon (l_m /r)^6$, between colloids and the membrane, where $\varepsilon$ is the strength of the adhesion energy and, $r$ and $l_m$ are, respectively, the center to center distance and the minimum allowed separation between colloids and membrane beads. Finally, we need to give the membrane an anisotropic shape for which we deform our spherical membrane into a prolate ellipsoid. In order to do so, we introduce two weak (compared to the strength of the adhesion energy) spring-like potentials between two small areas of the vesicle (the two poles of the ellipsoid) and the center of the vesicle, $u_{\mathrm{Ell}}= K_{\mathrm{Ell}} (L-a)^2$; $K_{\mathrm{Ell}} $, $a$ and $L$ are the potential strength, the major axis of the ellipsoid and the length of any line connecting the beads situated at the poles of the ellipsoid to the center, respectively. Since the adhesion energy is stronger than the applied harmonic potential, colloids effectively do not feel any difference between the energy cost for bending the membrane at these two areas and at the regions belonging to the rest of the ellipsoid. We verified this claim by considering a spherical membrane, and find that there is no significant difference between the case of including $u_{\mathrm{Ell}}$ with $a$ being the radius of the vesicle, and the case we do not include such a potential. \\
	Having defined all the contributions to the total energy of the membrane ($u_{\mathrm{Total}} = u_\mathrm{Curv} + u_{\mathrm{A}} + u_{\mathrm{V}} + u_{\mathrm{Ad}} +  u_{\mathrm{Ell}}$), we perform Monte Carlo simulations to reach the equilibrium shape of an ellipsoid containing an arbitrary number of colloids. To do so, we implement the Metropolis algorithm, in which we have three types of moves: we can modify the position of a random bead of the membrane, impose a rearrangement in the connections of beads, or move the colloids around. The first two moves are energetically evaluated based on the total energy, while any changes in the position of colloids are only based on the adhesion energy.\\
	During the simulations, we keep the number of particles constant and set all the relevant parameters as: $\kappa = 36 k_{\mathrm{B}}T$, $\varepsilon  = 8.5 k_{\mathrm{B}}T$,  $K_{\mathrm{A}} = 2\times 10^3 k_{\mathrm{B}}T/\sigma^2$, $ K_{\mathrm{V}} = 250 k_{\mathrm{B}}T/\sigma^3$ and $K_{\mathrm{Ell}} = 0.1 k_{\mathrm{B}}T/\sigma^2$, where $k_{\mathrm{B}}T$ is the thermal energy and $\sigma$ is the diameter of the beads constructing the membrane. The diameter of the colloids is set to $\sigma_{\mathrm{Coll}} = 5\sigma$.
	\section{Results and discussion}	
	First, we analyze the interaction between two colloids adhered to the surface of two vesicles of different sizes. We keep the size of colloids and beads the same in both cases.  We use umbrella sampling \cite{torrie1977nonphysical} to calculate the excess energy of the membrane as a function of the distance between the colloids. In effect, we apply a harmonic potential $u = \frac{1}{2}k(D-D_0)^2$, as our biased potential, between the two colloids directed along the coordinate of interest in order to restrain the system to sample around each distance $D_0$. Having performed the sampling process, we use the weighted histogram analysis method (WHAM) for obtaining the optimal estimate of the unbiased probability distribution, from which we can calculate the free energy of the system. The free energy is calculated with respect to the initial position of the colloids $\Delta E = E (\text{at the coordinate of interest}) - E (\text{initial coordinate})$. The excess area of the membrane available for colloids to adhere to is equal in both vesicles. As illustrated in Fig. \ref{fig:TwoVesicles}, the depth of the excess energy of the membrane with a smaller radius is significantly larger. In contrast, for the larger vesicle after a short distance colloids do not feel each other and the energy becomes flat. As the only difference between two test cases is the curvature, we conclude that this effect is due to vesicles being of different radii.
	
	Next, we examine the interaction between two colloids on the surface of a quasi-ellipsoidal membrane. We position the colloids symmetrically along the major axis of the ellipsoid (see Fig. \ref{fig:fig2}d(\textit{i}) ). We repeat the sampling procedure for different aspect ratios, $e = a/b$, of the ellipsoid. Since the volume is conserved during the shape evolution, one can easily calculate the semi-minor axis, $b$, as: $b = \sqrt{3 V/4\pi a}$. As depicted in Fig. \ref{fig:fig2}a, along the major axis colloids attract each other in order to minimize both the adhesion and curvature energies. Decreasing the asphericity of the ellipsoid ($e\rightarrow 1.0^ + $) in this direction enhances the deformable area (i.e. the number of accessible beads), hence the strength of the attraction energy increases.
		\begin{figure*}[ht]
			\centering
			\includegraphics[width=0.5\textwidth]{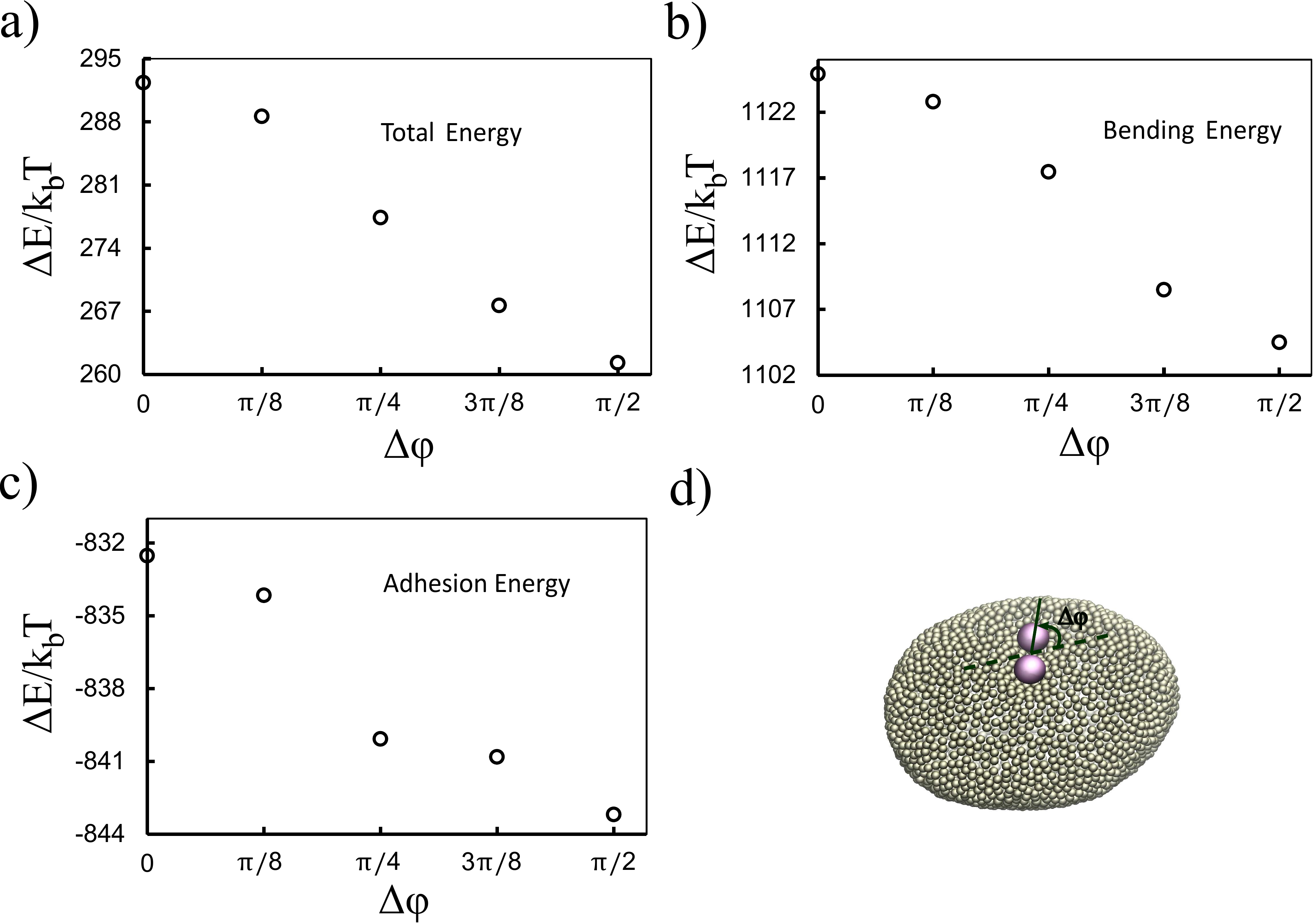}
			\caption{Bending is the dominant term in the attraction of colloids. As shown in (b) and (d), both the adhesion energy and the curvature energy are decreasing when the pair of colloids gets aligned with semi-minor axis of the ellipsoid. The latter has a larger contribution in the total energy.}
			\label{fig:fig3}
		\end{figure*}
	Similarly, particles that are situated along the semi-minor axis (as depicted in Fig. \ref{fig:fig2}d(\textit{ii})) attract each other. There is, however, an important difference between the two directions. In contrast to the previous case, decreasing the asphericity of the ellipsoid makes the attraction force between colloids weaker. Since the number of membrane beads adhered to each colloid remains the same, this behavior cannot be explained by the adhesion energy of the membrane. 

\begin{figure*}[ht]
			\centering
			\includegraphics[width=0.42\textwidth]{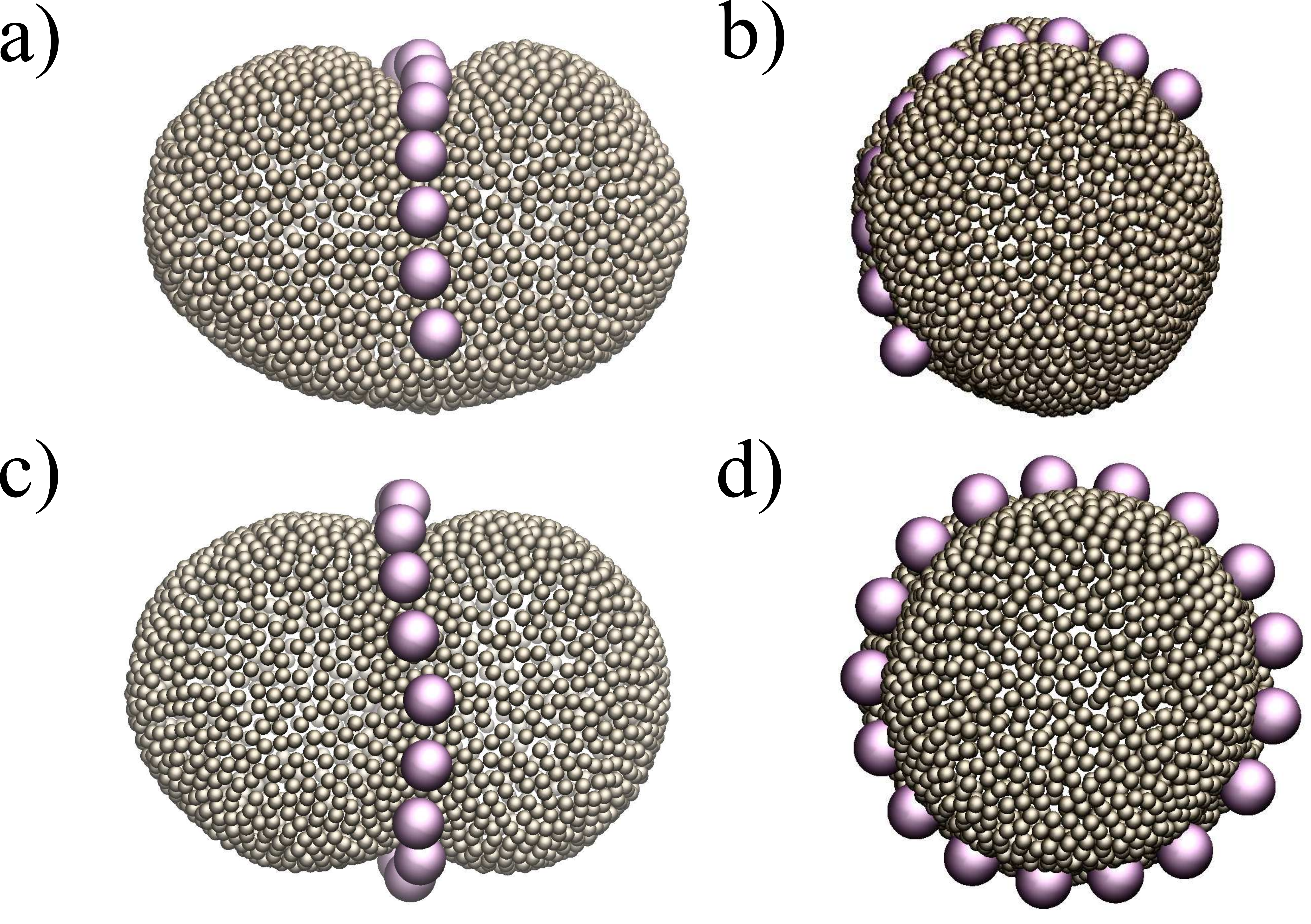}
			\caption{Colloids attract each other on ellipsoids, and in order to minimize the curvature energy, they form an arc (a $\&$ b) and a ring (c $\&$ d) at the mid-plane of the ellipsoid.}
			\label{fig:fig4}
		\end{figure*}

\begin{figure*}[ht]
			\centering
			\includegraphics[width=0.48\textwidth]{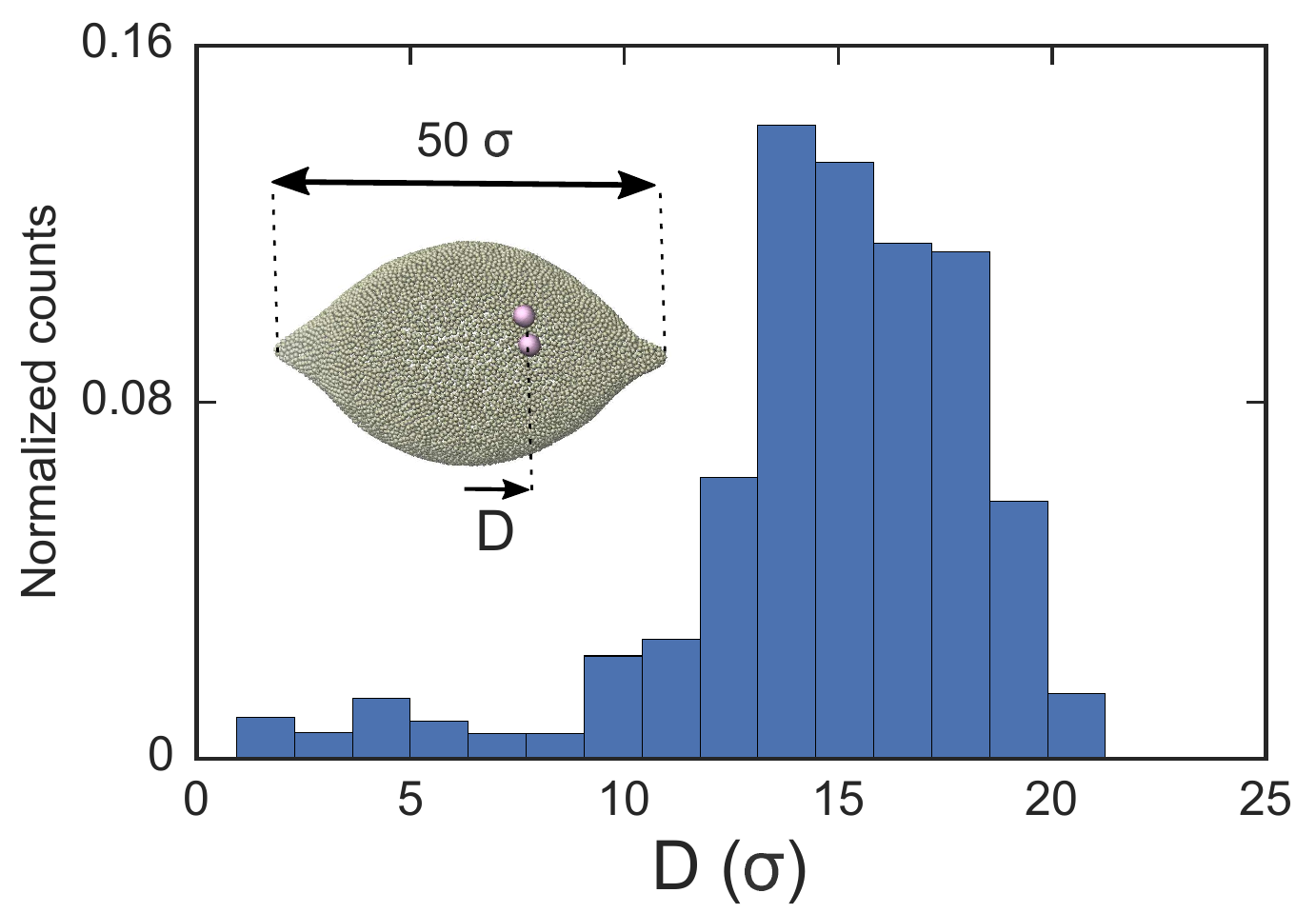}
			\caption{Histogram of the geometric center of a colloid dimer on a vesicle (of size $D_v = 40\sigma$) with negatively curved regions (inset). The dimer spends most of the time in the negatively curved part of the vesicle.}
			\label{fig:fig5}
		\end{figure*}

	To illuminate the reason that colloids select the direction along the minor axis to attract each other, we investigate the energy of a pair of colloids along a different coordinate. As shown in Fig. \ref{fig:fig2}d(\textit{iii}), we rotate a pair of colloids, that are constrained at a fixed distance to their center, along the angle spanning the space between the semi-major and -minor axes. As Fig. \ref{fig:fig2}c depicts, the most energetically favorable configuration is when the colloids are aligned with the direction perpendicular to the major axis. This itself introduces a mechanism by which, without involving any other factors, two colloids find the mid-plane perpendicular to the symmetry axes of the ellipsoid, as it minimizes the total energy of the membrane. In contrast, in the case of having a perfect spherical membrane, it is not possible to predict the localization of colloidal aggregates as it will be randomly chosen. Increasing the major axis of the membrane (making $ e $ larger), drives the colloid reorientation stronger. One should be careful about the values for the bending moduli and adhesion coefficients during the simulations, as it can cause an effect where colloids are arrested and prevented from diffusing on the surface of the membrane \cite{vsaric2012fluid}. In addition, a very high value of $ K_{\mathrm{Ell}} $, in addition to influencing the adhesion energy between the colloids and the membrane, would also pull two tubes out of the vesicle.\\
	Although the above results quantitatively show different behavior in two directions, the dominant contribution in the total energy of the membrane causing this effect is not yet clarified. In order to approximately determine it, we proceed as follows: we pick a vesicle with $e= 1.351$ and constrain the position of the colloids with a strong potential. Here, in contrast to earlier, we do not use the sampling method. Instead we let the system explore possible configurations of the membrane after reaching equilibrium, and then take the average of the energies for all those configurations. As depicted in Fig. \ref{fig:fig3}, both the adhesion energy and the curvature of the membrane decrease when the angle between the line connecting two colloids and the semi-major axis of the ellipsoid (Fig. \ref{fig:fig3}d) approaches $\pi /2 $. The bending energy, as quantified in Fig. \ref{fig:fig3}b, has a larger contribution to the total energy than the adhesion energy (Fig. \ref{fig:fig3}c).\\
	Putting all the results together, we expect that when we have more than two colloids they will initially attract each other to form linear aggregates (to minimize the adhesion energy), and afterwards these aggregations change their orientation to align with the minor axes of the ellipsoid. This is indeed what we observe in our simulations. Fig. \ref{fig:fig4} depicts the equilibrium shape of the membrane for different numbers of colloids. In all the test cases colloids tend to form a ring-like structure in the mid-plane of the ellipsoid. With a sufficiently large number of colloids (Figs. \ref{fig:fig4}c and \ref{fig:fig4}d), they form a full ring in this plane (see also the supplemental movie SM1). It is important to mention that these patterns are quite stable during the whole simulation. In contrast, in spherical vesicles there is no preferred direction for the aggregation of particles. Although colloids attract each other on a spherical membrane (Fig. \ref{fig:TwoVesicles}), there is no preference for the direction of the attraction. This means that even in case of forming a perfect ring on a vesicle, particles self-assemble in an arbitrary direction on the membrane. 
	\\
	As the final experiment, we look at the movement of particles on a vesicle having negatively curved regions. To do so, we first overstretch the springs (by which we give quasi-ellipsoidal shape to vesicles) and form negatively curved regions in a big vesicle ($D_v = 40 \sigma$). Having inserted a dimer in the system, we then look at the migration of the dimer. As shown in Fig. \ref{fig:fig5} in this case the dimer does not stay at the mid-plane of the vesicle. It instead spends much of its time during MC simulation at the regions that are negatively curved. Since the springs are overstretched, in the regions close to poles there is no excess area for the dimer to adhere to and therefore the dimer cannot explore that area (see also supplemental movies SM2-4). This type of dimer migration toward the areas having higher deviatoric curvature is also experimentally observed in Ref. \cite{li2016curvature}.\\

	The type of pattern formation we observe in our simulations is reminiscent of recruiting proteins by the membrane during different biological processes. It has been shown that, for example, dynamin proteins form a ring like structure during exocytosis to facilitate membrane scission \cite{pucadyil2008real} and that FtsZ proteins self-assemble into rings during the last step of bacterial cell division, namely cytokinesis \cite{shlomovitz2009membrane}. Because most of the proteins in biological cells are either anchored to or embedded in the membrane, their interaction is a response to the deformation of the membrane they themselves impose. As in our simulation the varying curvature is a determining factor that drives the pattern formation, we can relate our results to those membrane trafficking machinery functions. Although in this study we adjusted the included harmonic potential strength $K_{\mathrm{Ell}}$ such that it would not affect the interaction of the colloids with the membrane, it has been proven that during the cell division we have the same situation. Cytoplasmic dynein, as a multi-subunit molecular motor, generates the force that is exploited by the cell to direct the orientation of the division axis by mitotic spindles \cite{lesman2014contractile}. Our results show that curvature inhomogeneity and anisotropy can at least facilitate the process of protein self-assembly in the mid-plane of the cell.\\
	Although we have only investigated the interaction between identical isotropic inclusions, our results can explain the behavior of a system containing anisotropically shaped inclusions as well. Based on the local deformation of an ellipsoid, we expect that anisotropic inclusions adhered to a spherical membrane attract each other in the direction of negative curvature (with respect to the curvature of the membrane). This situation corresponds to having an isotropic inclusion embedded in a membrane with an anisotropic shape, which is the case we have studied here. 
	
	\section{Conclusion}	
	We studied the role of curvature heterogeneity and anisotropy on the interaction between colloids adhered to a membrane. First, we showed that the strength of the interaction between two colloids on the surface of a spherical vesicle is altered by changing the size of the vesicle. Next, we focused on such interactions on a membrane with an ellipsoidal shape. We revealed that the interaction on such an inhomogeneously shaped membrane depends on direction. For example, decreasing the asphericity of an ellipsoidal membrane makes the attraction between the colloids stronger along the semi-major axis and weaker in the semi-minor direction. Similarly, it has been previously shown in simulations that, on an elastic cylindrical membrane, colloids assemble perpendicularly to its major axis in the regime dominated by the bending energy \cite{pamies2011reshaping,vsaric2010effective}. In case of fluid membranes, through an analytical framework, it has also been shown that inclusions \textit{``embedded''} in a tubular membrane can attract each other in a transversal direction\cite{Vahid2016Point}. Simulating a vesicle containing many colloids, we showed how they form a ringlike structure around the mid-plane of the ellipsoid. While the cluster of colloids freely explores all the surface of a spherical membrane, less curved area energetically is more favorable for colloids on an ellipsoid. Our results suggest that forming regions of different curvatures on membrane vesicles can control pattern formation of inclusions, and this can be important from both nanotechnological application and biological points of view.
	\section*{Acknowledgements}
	This work was supported by the Netherlands Organisation for Scientific Research (NWO/OCW), as part of the Frontiers of Nanoscience program.
	\footnotesize{
		\providecommand*{\mcitethebibliography}{\thebibliography}
		\csname @ifundefined\endcsname{endmcitethebibliography}
		{\let\endmcitethebibliography\endthebibliography}{}

	}
	
\end{document}